\documentclass[12pt]{article}
\usepackage{latexsym,amssymb,amsmath}
\begin{document}
\baselineskip=20pt

\newcommand{\la}{\langle}
\newcommand{\ra}{\rangle}
\newcommand{\psp}{\vspace{0.4cm}}
\newcommand{\pse}{\vspace{0.2cm}}
\newcommand{\ptl}{\partial}
\newcommand{\dlt}{\delta}
\newcommand{\sgm}{\sigma}
\newcommand{\al}{\alpha}
\newcommand{\be}{\beta}
\newcommand{\G}{\Gamma}
\newcommand{\gm}{\gamma}
\newcommand{\vs}{\varsigma}
\newcommand{\Lmd}{\Lambda}
\newcommand{\lmd}{\lambda}
\newcommand{\td}{\tilde}
\newcommand{\vf}{\varphi}
\newcommand{\yt}{Y^{\nu}}
\newcommand{\wt}{\mbox{wt}\:}
\newcommand{\rd}{\mbox{Res}}
\newcommand{\ad}{\mbox{ad}}
\newcommand{\stl}{\stackrel}
\newcommand{\ol}{\overline}
\newcommand{\ul}{\underline}
\newcommand{\es}{\epsilon}
\newcommand{\dmd}{\diamond}
\newcommand{\clt}{\clubsuit}
\newcommand{\vt}{\vartheta}
\newcommand{\ves}{\varepsilon}
\newcommand{\dg}{\dagger}
\newcommand{\tr}{\mbox{Tr}}
\newcommand{\ga}{{\cal G}({\cal A})}
\newcommand{\hga}{\hat{\cal G}({\cal A})}
\newcommand{\Edo}{\mbox{End}\:}
\newcommand{\for}{\mbox{for}}
\newcommand{\kn}{\mbox{ker}}
\newcommand{\Dlt}{\Delta}
\newcommand{\rad}{\mbox{Rad}}
\newcommand{\rta}{\rightarrow}
\newcommand{\mbb}{\mathbb}
\newcommand{\lra}{\Longrightarrow}
\newcommand{\X}{{\cal X}}
\newcommand{\Y}{{\cal Y}}
\newcommand{\Z}{{\cal Z}}
\newcommand{\U}{{\cal U}}
\newcommand{\V}{{\cal V}}
\newcommand{\W}{{\cal W}}
\newcommand{\sech}{\mbox{sech}\:}
\newcommand{\csch}{\mbox{csch}\:}
\newcommand{\sn}{\mbox{sn}\:}
\newcommand{\cn}{\mbox{cn}\:}
\newcommand{\dn}{\mbox{dn}\:}
\newcommand{\sta}{\theta}

\begin{center}{\Large \bf Asymmetric and Moving-Frame  Approaches to
}
\end{center}
\begin{center}{\Large \bf the 2D and 3D  Boussinesq Equations}
\end{center}

\vspace{0.2cm}

\begin{center}{\large Xiaoping Xu}\end{center}
\begin{center}{Institute of Mathematics, Academy of Mathematics \& System Sciences}\end{center}
\begin{center}{Chinese Academy of Sciences, Beijing 100190, P.R. China}
\footnote{Research supported
 by China NSF 10431040}\end{center}

\begin{center}{\it Dedicated to 2008 Beijing Olympic
Games}\end{center}

\vspace{0.3cm}

 \begin{center}{\Large\bf Abstract}\end{center}

\vspace{1cm} {\small Boussinesq systems of nonlinear partial
differential equations are fundamental equations in geophysical
fluid dynamics. In this paper, we  use asymmetric ideas  and moving
frames  to solve the two-dimensional Boussinesq equations with
partial viscosity terms studied by Chae ({\it Adv. Math.} {\bf 203}
(2006), 497-513) and the three-dimensional stratified rotating
Boussinesq equations studied by Hsia, Ma and Wang ({\it J. Math.
Phys.} {\bf 48} (2007), no. 6, 06560). We obtain new families of
explicit exact solutions with multiple parameter functions. Many of
them are the periodic, quasi-periodic, aperiodic solutions that may
have practical significance. By Fourier expansion and some of our
solutions, one can obtain discontinuous solutions. In addition, Lie
point symmetries are used to simplify our arguments.}

\section{Introduction}

Both the atmospheric and oceanic flows are  influenced by the
rotation of the earth. In fact, the fast rotation and small aspect
ratio are two main characteristics of the large scale atmospheric
and oceanic flows. The small aspect ratio characteristic leads to
the primitive equations, and the fast rotation leads to the
quasi-geostropic equations (cf. [2], [6], [7], [9]). A main
objective in climate dynamics and in geophysical fluid dynamics is
to understand and predict the periodic, quasi-periodic, aperiodic,
and fully turbulent characteristics of the large scale atmospheric
and oceanic flows (e.g., cf. [4], [5]).

The  Boussinesq system for the incompressible fluid follows in
$\mbb{R}^2$ is
$$u_t+uu_x+vu_y-\nu\Dlt u=-p_x,\qquad v_t+uv_x+vv_y-\nu\Dlt
v-\sta=-p_y,\eqno(1.1)$$
$$\sta_t+u\sta_x+v\sta_y-\kappa \Dlt\sta=0,\qquad
u_x+v_y=0,\eqno(1.2)$$ where $(u,v)$ is the velocity vector field,
$p$ is the scalar pressure, $\sta$ is the scalar temperature,
$\nu\geq 0$ is the viscosity and $\kappa\geq 0$ is the thermal
diffusivity. The above system is a simple model in atmospheric
sciences (e.g., cf. [8]). Chae [1] proved the global regularity, and
Hou and Li [3] obtained the well-posedness of the above system.

 Aonther slightly simplified version of the
system of  primitive equations is the three-dimensional stratified
rotating Boussinesq system (e.g., cf. [7], [9]):
$$u_t+uu_x+vu_y+wu_z-\frac{1}{R_0}v=\sgm(\Dlt u-p_x),\eqno(1.3)$$
$$v_t+uv_x+vv_y+wv_z+\frac{1}{R_0}u=\sgm(\Dlt v-p_y),\eqno(1.4)$$
$$w_t+uw_x+vw_y+ww_z-\sgm R T=\sgm(\Dlt w-p_z),\eqno(1.5)$$
$$T_t+uT_x+vT_y+wT_z=\Dlt T+w,\eqno(1.6)$$
$$ u_x+v_y+w_z=0,\eqno(1.7)$$
where $(u,v,w)$ is the velocity vector filed, $T$ is the temperature
function, $p$ is the pressure function, $\sgm$ is the Prandtle
number, $R$ is the thermal Rayleigh number and $R_0$ is the Rossby
number. Moreover, the vector $(1/R_0)(-v,u,0)$ represents the
Coriolis force and the term $w$ in (1.6) is derived using
stratification. So the above equations are the extensions of
Navier-Stokes equations by adding the Coriolis force and the
stratified temperature equation. Due to the Coriolis force, the
two-dimensional system (1.1) and (1.2) is not a special case of the
above three-dimensional system. Hsia, Ma and Wang [4] studied the
bifurcation and periodic solutions of the above system (1.3)-(1.7).

In [10], we used the stable range of nonlinear term to solve the
equation of nonstationary transonic gas flow. Moreover, we [11]
solved the three-dimensional Navior-Stokes equations by asymmetric
techniques and moving frames. Based on the algebraic characteristics
of the above equations, we use in this paper asymmetric ideas and
moving frames to solve the above two Boussinesq systems of partial
differential equations. New families of explicit exact solutions
with multiple parameter functions are obtained. Many of them are the
periodic, quasi-periodic, aperiodic solutions that may have
practical significance. Using Fourier expansion and some of our
solutions, one can obtain discontinuous solutions. The symmetry
transformations for these equations are used to simplify our
arguments.

For convenience, we always assume that all the involved partial
derivatives of related functions always exist and we can change
orders of taking partial derivatives. The parameter functions are so
chosen that the involved expressions make sense. We also use prime
$'$ to denote the derivative of any one-variable function.

Observe that the two-dimensional Boussinesq system (1.1) and (1.2)
is invariant under the action of the following symmetry
transformation:
$${\cal T}(u)=a^{-1}\es_1u(a^2(t+b),a\es_1(x+\al),a\es_2(y+\be))-\al',\eqno(1.8)$$
$${\cal T}(v)=a^{-1}\es_2v(a^2(t+b),a(x+\al),a(y+\be))-\be',\eqno(1.9)$$
$${\cal T}(p)=a^{-2}p(a^2(t+b),a\es_1(x+\al),a\es_2(y+\be))+{\al'}'x+{\be'}'y+\gm,\eqno(1.10)$$
$${\cal T}(\sta)=a^{-3}\es_2\sta(a^2(t+b),a\es_1(x+\al),a\es_2(y+\be)),\eqno(1.11)$$
where $a,b\in\mbb{R}$ with $a\neq 0$, $\es_1,\es_2\in\{1,-1\}$ and
$\al,\be,\gm$ are arbitrary functions of $t$. The above
transformation transforms a solution of the equation (1.1) and (1.2)
into another solution with additional three parameter functions.

 Denote $\vec x=(x,y)$. The three-dimensional stratified rotating Boussinesq system
 is invariant under the following transformations:
$${\cal T}_1[(u,v,w)]=((u(t+b,\vec x A,\es z),v(t+b,\vec x
A),\es z)A,\es w),\eqno(1.12)$$
$${\cal T}_1(p)=p(t+b,\vec x A,\es z),\qquad
{\cal T}_1(T)=T(t+b,\vec x A,\es z);\eqno(1.13)$$
$${\cal T}_2(u)=u(t,x+\al,y+\be,z+\gm)-\al',\qquad
{\cal T}_2(v)=v(t,x+\al,y+\be,z+\gm)-\be',\eqno(1.14)$$
$${\cal T}_2(w)=w(t,x+\al,y+\be,z+\gm)-\gm',\qquad
{\cal T}_2(T)=T(t,x+\al,y+\be,z+\gm)-\gm,\eqno(1.15)$$
$${\cal
T}_2(p)=p(t,x+\al,y+\be,z+\gm)+\sgm^{-1}({\al'}'x+{\be'}'y+{\gm'}'z)-R\gm
z+\mu;\eqno(1.16)$$ where $\es=\pm 1$, $b\in\mbb{R}$, $A\in
O(2,\mbb{R})$, and $\al,\be,\gm,\mu$ are arbitrary functions of $t$.
The above transformations transform a solution of the equation
(1.3)-(1.7) into another solution. In particular, applying the
transformation ${\cal T}_2$ to any solution in this paper yields
another solution with extra four parameter functions.

To simplify problems, we always solve the Boussinesq systems modulo
the above corresponding symmetry transformations, which is an idea
that geometers and topologists often use.

The paper is organized as follows. In Section 2, we solve the
two-dimensional Boussinesq equations (1.1)-(1.2) and obtain four
families of explicit exact solutions. In Section 3, we present an
approach with $u,v,w,T$ linear in $x,y$ to the equations
(1.3)-(1.7), and obtain two families of explicit exact solutions.
Assuming $u_z=v_z=w_{zz}=T_{zz}=0$ in Section 4, we find another two
families of explicit exact solutions of the equations (1.3)-(1.7).
In Section 5, we obtain  a family of explicit exact solutions of
(1.3)-(1.7) that are independent of $x$. The status can be changed
by applying the transformation in (1.12) and (1.13) to them.

\section{Solutions of the 2D Boussinesq Equations}

In this section, we solve the two-dimensional Boussinesq equations
(1.1)-(1.2) by an asymmetric method and by an moving frame.

According to the second equation in (1.2), we take the potential
form:
$$u=\xi_y,\qquad v=-\xi_x\eqno(2.1)$$
for some functions $\xi$ in $t,x,y$. Then the two-dimensional
Boussinesq equations become
$$\xi_{yt}+\xi_y\xi_{xy}-\xi_x\xi_{yy}-\nu\Dlt \xi_y=-p_x,\qquad \xi_{xt}+\xi_y\xi_{xx}-\xi_x\xi_{xy}-\nu\Dlt
\xi_x+\sta=p_y,\eqno(2.2)$$
$$\sta_t+\xi_y\sta_x-\xi_x\sta_y-\kappa \Dlt\sta=0.\eqno(2.3)$$
By our assumption $p_{xy}=p_{yx}$, the compatible condition of the
equations in (2.2) is
$$(\Dlt \xi)_t+\xi_y(\Dlt \xi)_x-\xi_x(\Dlt
\xi)_y-\nu\Dlt^2\xi+\sta_x=0.\eqno(2.4)$$ Now we first solve the
system (2.3) and (2.4).

Our  asymmetric approach is to assume
$$\sta=\ves(t,y),\qquad\xi=\phi(t,y)+x\psi(t,y)\eqno(2.5)$$
for some functions $\ves,\phi$ and $\psi$ in $t,y$. Then (2.3)
becomes
$$\ves_t-\psi\ves_y-\kappa\ves_{yy}=0.\eqno(2.6)$$
Moreover, (2.4) becomes
$$\phi_{yyt}+x\psi_{yyt}+(\phi_y+x\psi_y)\psi_{yy}-\psi(\phi_{yyy}+x\psi_{yyy})-\nu(\phi_{yyyy}+x\psi_{yyyy})=0,
\eqno(2.7)$$ equivalently,
$$\phi_{yyt}+\phi_y\psi_{yy}-\psi\phi_{yyy}-\nu\phi_{yyyy}=0,
\eqno(2.8)$$
$$\psi_{yyt}+\psi_y\psi_{yy}-\psi\psi_{yyy}-\nu\psi_{yyyy}=0.
\eqno(2.9)$$ The above two equations are equivalent to:
$$\phi_{yt}+\phi_y\psi_y-\psi\phi_{yy}-\nu\phi_{yyy}=\al_1,
\eqno(2.10)$$
$$\psi_{yt}+\psi_y^2-\psi\psi_{yy}-\nu\psi_{yyy}=\al_2
\eqno(2.11)$$ for some functions $\al_1$ and $\al_2$ of $t$ to be
determined.

Let $c$ be a fixed real constant and let $\gm$ be a fixed function
of $t$. We define
$$\zeta_1(s)=\frac{e^{\gm s}-ce^{-\gm s}}{2},\qquad \eta_1=\frac{e^{\gm s}+ce^{-\gm
s}}{2},\eqno(2.12)$$
$$\zeta_0(s)=\sin\gm s,\qquad \eta_0(s)=\cos\gm s.\eqno(2.13)$$
Then
$$\eta_r^2(s)+(-1)^r\zeta_r^2(s)=c^r\eqno(2.14)$$
and
$$\ptl_s(\zeta_r(s))=\gm\eta_r(s),\qquad
\ptl_s(\eta_r(s))=-(-1)^r\gm\zeta_r(s),\eqno(2.15)$$ where we treat
$0^0=1$ when $c=r=0$. First we assume
$$\psi=\be_1y+\be_2\zeta_r(y)\eqno(2.16)$$
for some functions $\be_1$ and $\be_2$ of $t$, where $r=0,1$. Then
 (2.11) becomes
\begin{eqnarray*}\hspace{2cm}& &\be_1'+c^r\be_2^2\gm^2+\be_1^2+[(\be_2\gm)'+(-1)^r\nu\be_2\gm^3+2\be_1\be_2\gm]\eta_r(y)
\\ & &+(-1)^r\be_2\gm(\be_1\gm-\gm')y\zeta_r(y)=\al_2,\hspace{6.2cm}
(2.17)\end{eqnarray*} which is implied by the following equations:
$$\be_1'+c^r\be_2^2\gm^2+\be_1^2=\al_2,\qquad\be_1\gm-\gm'=0,\eqno(2.18)$$
$$(\be_2\gm)'+(-1)^r\nu\be_2\gm^3+2\be_1\be_2\gm=0.\eqno(2.19)$$
For convenience, we assume $\gm=\sqrt{\al'}$ for some function $\al$
of $t$. Thus we have
$$\be_1=\frac{\gm'}{\gm}=\frac{{\al'}'}{2\al'},\qquad
\be_2=\frac{b_1e^{-(-1)^r\nu\al}}{\sqrt{(\al')^3}},\qquad
b_1\in\mbb{R}.\eqno(2.20)$$ To solve (2.10), we assume
$$\phi=\be_3\eta_r(y)\eqno(2.21)$$
 for some function $\be_3$, modulo the transformation in
(1.8)-(1.11).
 Now (2.10) becomes
$$[(-1)^r((\be_3\gm)'+\be_1\be_3\gm)+\nu\be_3\gm^3]\zeta_r(y)=-\al_1, \eqno(2.22)$$ which is
implied by
$$(-1)^r((\be_3\gm)'+\be_1\be_3\gm)+\nu\be_3\gm^3=0.\eqno(2.23)$$ Thus
$$\be_3=\frac{b_2e^{-(-1)^r\nu\al}}{\al'},\eqno(2.24)$$
 where $b_2$ is a real constant.

In order to solve (2.6), we assume
$$\ves=be^{\gm_1\eta_r(y)},\eqno(2.25)$$ where $b$ is a real constant and $\gm_1$ is a function of $t$.
Then (2.6) is implied by
$$\gm_1'\eta_r(y)+(-1)^r\be_2\gm\gm_1\zeta_r^2(y)+\kappa\gm^2\gm_1((-1)^r\eta_r(y)-\gm_1\zeta_r^2(y))=0,
\eqno(2.26)$$ which is implied by
$$\gm_1'+(-1)^r\kappa\gm^2\gm_1=0,\qquad(-1)^r\be_2-\kappa\gm\gm_1=0.\eqno(2.27)$$
Then the first  equation implies
$$\gm_1=b_3e^{-(-1)^r\kappa\al}\eqno(2.28)$$
for some constant $b_3$. By the second equations in (2.20) and
(2.27), we have:
$$(-1)^r\frac{b_1e^{-(-1)^r\nu\al}}{\sqrt{(\al')^3}}=b_3\kappa\sqrt{\al'}e^{-(-1)^r\kappa\al}.\eqno(2.29)$$
For convenience, we take
$$b_1=(-1)^rb_0^2\kappa b_3,\qquad b_0\in\mbb{R}.\eqno(2.30)$$
Then (2.29) is implied by
$$\al'e^{(-1)^r(\nu-\kappa)\al/2}=b_0.\eqno(2.31)$$
If $\nu=\kappa$, then we have $\al=b_0t+c_0$. Modulo the
transformation in (1.8)-(1.11), we take $b_0=1$ and $c_0=0$, that
is, $\al=t$.
 When $\nu\neq
\kappa$, we similarly take $b_0=1$ and
$$\al=\frac{2(-1)^r}{\nu-\kappa}\ln[(-1)^r(\nu-\kappa)t/2+c_0],\qquad c_0\in\mbb{R}.\eqno(2.32)$$

Suppose $\nu=\kappa$. Then $\gm=1$ and
$$\phi=b_2e^{-(-1)^r\nu t}\eta_r(y),\qquad\psi=(-1)^rb_3\nu e^{-(-1)^r\nu
t}\zeta_r(y).\eqno(2.33)$$ Moreover,
$$\sta=b\exp(b_3e^{-(-1)^r\nu t}\eta_r(y)),\eqno(2.34)$$
$$\xi=b_2e^{-(-1)^r\nu t}\eta_r(y)+(-1)^rb_3\nu e^{-(-1)^r\nu
t}x\zeta_r(y)\eqno(2.35)$$ by (2.5). According to (2.1),
$$u=\xi_y=(-1)^r[-b_2e^{-(-1)^r\nu t}\zeta_r(y)+b_3\nu e^{-(-1)^r\nu
t}x\eta_r(y)],\eqno(2.36)$$ $$ v=-\xi_x=-(-1)^rb_3\nu e^{-(-1)^r\nu
t}\zeta_r(y).\eqno(2.37)$$ Note $$u_t+uu_x+vu_y-\nu\Dlt u=
b_3^2\nu^2c^r e^{-(-1)^r2\nu t}x,\eqno(2.38)$$
$$v_t+uv_x+vv_y-\nu\Dlt v-\sta
=vv_y-b\exp(b_3e^{-(-1)^r\nu t}\eta_r(y)).\eqno(2.39)$$ By (1.1),
we have
$$p=
b\int\exp(b_3e^{-(-1)^r\nu t}\eta_r(y))dy-\frac{1}{2}b_3^2\nu^2
e^{-(-1)^r2\nu t}(c^rx^2+\zeta_r^2(y))\eqno(2.40)$$ modulo the
transformation in (1.8)-(1.11).

Consider the case $\nu\neq \kappa$. Then
$$\gm=\sqrt{\al'}=\frac{1}{\sqrt{(-1)^r(\nu-\kappa)t/2+c_0}}\eqno(2.41)$$
by (2.32). Moreover,
$$\phi=b_2[(-1)^r(\nu-\kappa)t/2+c_0]^{2\nu/(\kappa-\nu)+1}\eta_r(y)\eqno(2.42)$$
by (2.21) and (2.24). Furthermore,
$$\psi=\frac{(-1)^r(\kappa-\nu)y}{4[(-1)^r(\nu-\kappa)t/2+c_0]}+
(-1)^rb_3\kappa
[(-1)^r(\nu-\kappa)t/2+c_0]^{2\nu/(\kappa-\nu)+3/2}\zeta_r(y)\eqno(2.43)$$
by (2.16), (2.20) and (2.30). According to (2.25), (2.28) and
(2.32),
$$\sta=be^{b_3[(-1)^r(\nu-\kappa)t/2+c_0]^{2\kappa/(\kappa-\nu)}\eta_r(y)}.\eqno(2.44)$$
Similarly, we have
\begin{eqnarray*}\hspace{1cm}u_t+uu_x+vu_y-\nu\Dlt
u&=&b_3^2c^r\kappa^2
[(-1)^r(\nu-\kappa)t/2+c_0]^{4\nu/(\kappa-\nu)+2}x
\\ & &+\frac{3(\nu-\kappa)^2x}{16[(-1)^r(\nu-\kappa)t/2+c_0]^2},
\hspace{3.8cm}(2.45)\end{eqnarray*}
\begin{eqnarray*}& &v_t+uv_x+vv_y-\nu\Dlt-\sta
=-\psi_t+\psi\psi_y+\nu\psi_{yy}-\sta\\
&=&-be^{b_3[(-1)^r(\nu-\kappa)t/2+c_0]^{2\kappa/(\kappa-\nu)}\eta_r(y)}
+\frac{3}{4}b_3\kappa(\kappa-\nu)
[(-1)^r(\nu-\kappa)t/2+c_0]^{2\nu/(\kappa-\nu)+1/2}\zeta_r(y)
\\ & &+\frac{3(\nu-\kappa)^2y}{16[(-1)^r(\nu-\kappa)t/2+c_0]^2}+
\frac{b_3^2}{2}\kappa^2
[(-1)^r(\nu-\kappa)t/2+c_0]^{4\nu/(\kappa-\nu)+3}\ptl_y\zeta_r^2(y).\hspace{0.6cm}(2.46)
\end{eqnarray*}
According (1.1), we have
\begin{eqnarray*}p&=&b\int
e^{b_3[(-1)^r(\nu-\kappa)t/2+c_0]^{2\kappa/(\kappa-\nu)}\eta_r(y)}dy
-\frac{b_3^2}{2}c^r\kappa^2
[(-1)^r(\nu-\kappa)t/2+c_0]^{4\nu/(\kappa-\nu)+2}x^2
\\ & &-\frac{3(\nu-\kappa)^2(x^2+y^2)}{32[(-1)^r(\nu-\kappa)t/2+c_0]^2}
-\frac{b_3^2}{2}\kappa^2
[(-1)^r(\nu-\kappa)t/2+c_0]^{4\nu/(\kappa-\nu)+3}\zeta_r^2(y)
\\
& &+\frac{3}{4}(-1)^rb_3\kappa(\kappa-\nu)
[(-1)^r(\nu-\kappa)t/2+c_0]^{2\nu/(\kappa-\nu)+1}\eta_r(y)\hspace{3.6cm}(2.47)\end{eqnarray*}
modulo the transformation in (1.8)-(1.11).\psp

{\bf Theorem 2.1}. {\it Let $b,b_2,b_3,c,c_0\in\mbb{R}$ and let
$r=0,1$. If $\nu=\kappa$,  we have the solution (2.34), (2.36),
(2.37) and (2.40)  of the two-dimensional Boussinesq equations
(1.1)-(1.2), where $\zeta_r(y)$ and $\eta_r(y)$ are defined in
(2.12)-(2.13) with $\gm=1$. When $\nu\neq\kappa$,  we have the
following solutions of the two-dimensional Boussinesq equations
(1.1)-(1.2):
\begin{eqnarray*}u&=&\frac{(-1)^r(\kappa-\nu)x}{4[(-1)^r(\nu-\kappa)t/2+c_0]}+
(-1)^rb_3\kappa
[(-1)^r(\nu-\kappa)t/2+c_0]^{2\nu/(\kappa-\nu)+1}x\eta_r(y)\\ &&
-(-1)^rb_2[(-1)^r(\nu-\kappa)t/2+c_0]^{2\nu/(\kappa-\nu)+1/2}\zeta_r(y),
\hspace{5cm}(2.48)\end{eqnarray*}
$$v=\frac{(-1)^r(\nu-\kappa)y}{4[(-1)^r(\nu-\kappa)t/2+c_0]}-
(-1)^rb_3\kappa
[(-1)^r(\nu-\kappa)t/2+c_0]^{2\nu/(\kappa-\nu)+3/2}\zeta_r(y),\eqno(2.49)$$
$\sta$ is given in (2.44) and $p$ is given in (2.47), where
$\zeta_r(y)$ and $\eta_r(y)$ are defined in (2.12)-(2.13) with
$\gm=[(-1)^r(\nu-\kappa)t/2+c_0]^{-1/2}$.}\psp

Observe that
$$\psi=6\nu y^{-1}\eqno(2.50)$$
is another solution of (2.11). In order to solve (2.10), we assume
$$\phi=\sum_{i=1}^\infty\gm_iy^i\eqno(2.51)$$
modulo the transformation in (1.8)-(1.11), where $\gm_i$ are
functions of $t$ to be determined. Now (2.10) becomes
$$-6\nu\gm_1y^{-2}-18\nu\gm_2y^{-1}+\sum_{i=1}^\infty[i\gm_i'-\nu(i+2)(i+3)(i+4)\gm_{i+2}]
y^{i-1}=\al_1,\eqno(2.52)$$ equivalently,
$$\gm_1=\gm_2=0,\qquad \al_1=-60\nu\gm_3,\eqno(2.53)$$
$$i\gm_i'-\nu(i+2)(i+3)(i+4)\gm_{i+2}=0,\qquad
i> 1.\eqno(2.54)$$ Thus
$$\gm_{2i+2}=\frac{2i\gm_{2i}'}{\nu(2i+2)(2i+3)(2i+4)}=0,\qquad
i\geq 1,\eqno(2.55)$$
$$\gm_{2i+3}=\frac{(2i+1)\gm_{2i+1}'}{\nu(2i+3)(2i+4)(2i+5)}=\frac{360\gm_3^{(i)}}{\nu^i(2i+2)(2i+5)!},\qquad
i\geq 1.\eqno(2.56)$$ Hence
$$\phi=360\sum_{i=0}^\infty
\frac{\al^{(i)}y^{2i+3}}{\nu^i(2i+3)(2i+5)!},\eqno(2.57)$$ where
$\al$ is an arbitrary function of $t$ such that the series
converges, say, a polynomial in $t$.

To solve (2.6), we also assume
$$\ves=\sum_{i=0}^\infty\be_i y^i,\eqno(2.58)$$
where $\be_i$ are functions of $t$. Then (2.6) becomes
$$6\nu\be_1y^{-1}+\sum_{i=0}^\infty[\be_i'+(i+2)(6\nu-(i+1)\kappa)\be_{i+2}]y^i=0,\eqno(2.58)$$
that is, $\be_1=0$ and
$$\be_i'-(i+2)(6\nu+(i+1)\kappa)\be_{i+2}=0,\qquad i\geq 0.\eqno(2.59)$$
Hence
$$\sta=\be+\sum_{i=1}^\infty\frac{\be^{(i)}y^{2i}}{2^ii!\prod_{r=1}^i(6\nu+(2r-1)\kappa)},\eqno(2.60)$$
where $\be$ is an arbitrary function of $t$ such that the series
converges, say, a polynomial in $t$. In this case,
$$u_t+uu_x+vu_y-\nu\Dlt u=-60\nu\al,\eqno(2.61)$$
$$v_t+uv_x+vv_y-\nu\Dlt-\sta
=-36\nu^2
y^{-3}-\be-\sum_{i=1}^\infty\frac{\be^{(i)}y^{2i}}{2^ii!\prod_{r=1}^i(6\nu+(2r-1)\kappa)}.
\eqno(2.62)$$ According (1.1), we have
$$p=60\nu\al x-18\nu^2 y^{-2}+\be y+
\sum_{i=1}^\infty\frac{\be^{(i)}y^{2i+1}}{2^ii!(2i+1)\prod_{r=1}^i(6\nu+(2r-1)\kappa)}\eqno(2.63)$$
modulo the transformation in (1.8)-(1.11).\psp

{\bf Theorem 2.2}. {\it We have the following solutions of the
two-dimensional Boussinesq equations (1.1)-(1.2):
$$u=360\sum_{i=0}^\infty
\frac{\al^{(i)}y^{2i+2}}{\nu^i(2i+5)!}-6\nu xy^{-2},\qquad v=-6\nu
y^{-1},\eqno(2.64)$$ $\sta$ is given in (2.60) and $p$ is given in
(2.63), where $\al $ and $\be$ are arbitrary functions of $t$ such
that the  related series converge, say,  polynomials in $t$.}\psp

 Let $\gm$ be a function of $t$.
Denote the  moving frame
$$\td\varpi=x\cos\gm+y\sin\gm,\qquad \hat\varpi=y\cos\gm-x\sin\gm.\eqno(2.65)$$
Then
$$\ptl_t(\td\varpi)=\gm'\hat\varpi,\qquad
\ptl_t(\hat\varpi)=-\gm'\td\varpi.\eqno(2.66)$$ Moreover,
$$\ptl_{\td\varpi}=\cos\gm\:\ptl_x+\sin\gm\:\ptl_y,\qquad
\ptl_{\hat\varpi}=-\sin\gm\:\ptl_x+\cos\gm\:\ptl_y.\eqno(2.67)$$ In
particular,
$$\Dlt=\ptl_x^2+\ptl_y^2=\ptl_{\td\varpi}^2+\ptl_{\hat\varpi}^2.\eqno(2.68)$$

We assume
$$\xi=\phi(t,\td\varpi)-\frac{\gm'}{2}(x^2+y^2)
,\qquad\sta=\psi(t,\td\varpi),\eqno(2.69)$$ where $\phi$ and $\psi$
are functions in $t,\td\varpi$. Then (2.3) becomes
$$\psi_t-\kappa\psi_{\td\varpi\td\varpi}=0\eqno(2.70)$$
and (2.4) becomes
$$-2{\gm'}'+\phi_{t\td\varpi\td\varpi}
-\nu\phi_{\td\varpi\td\varpi\td\varpi\td\varpi}+\psi_{\td\varpi}\cos\gm
=0.\eqno(2.71)$$ Modulo the transformation in (1.8)-(1.11), the
above equation is equivalent to
$$-2{\gm'}'\td\varpi+\phi_{t\td\varpi}
-\nu\phi_{\td\varpi\td\varpi\td\varpi}+\psi\cos\gm =0.\eqno(2.72)$$

Assume $\nu=\kappa$. We take the following solution of (2.70):
$$\psi=\sum_{i=1}^m a_id_ie^{a_i^2\kappa t\cos 2b_i+a_i\td\varpi\cos
b_i}\sin(a_i^2\kappa t\sin 2b_i+a_i\td\varpi\sin
b_i+b_i+c_i)\eqno(2.73)$$ where $a_i,b_i,c_i,d_i$ are real numbers.
Moreover,  (2.72) is equivalent to solving the following equation:
\begin{eqnarray*}\hspace{1.5cm}& &2\nu\gm'-{\gm'}'\td\varpi^2+\phi_t
-\nu\phi_{\td\varpi\td\varpi}+[\sum_{i=1}^md_ie^{a_i^2\kappa t\cos
2b_i+a_i\td\varpi\cos b_i}\\ & &\times\sin(a_i^2\kappa t\sin
2b_i+a_i\td\varpi\sin
b_i+c_i)]\cos\gm=0\hspace{4.9cm}(2.74)\end{eqnarray*} by (2.1). Thus
we have the following solution of (2.74):
\begin{eqnarray*}\phi&=&-[\sum_{i=1}^md_ie^{a_i^2\kappa
t\cos 2b_i+a_i\td\varpi\cos b_i}\sin(a_i^2\kappa t\sin
2b_i+a_i\td\varpi\sin b_i+c_i)]\int \cos\gm\:dt\\ &
&+\gm'\td\varpi^2+\sum_{s=1}^n\hat d_se^{\hat a_s^2\kappa t\cos
2\hat b_s+\hat a_s\td\varpi\cos \hat b_s}\sin(\hat a_s^2\kappa t\sin
2\hat b_s+\hat a_s\td\varpi\sin \hat b_s+\hat
c_s),\hspace{1.6cm}(2.75)\end{eqnarray*} where $\hat a_s,\hat
b_s,\hat c_s,\hat d_s$ are real numbers.

Suppose $\nu\neq \kappa$. To make (2.72) solvable, we choose the
following solution of (2.70):
$$\psi=\sum_{i=1}^m a_id_ie^{a_i^2\kappa t+a_i\td\varpi}.\eqno(2.76)$$
Now  (2.72) is equivalent to solving the following equation:
$$\nu\gm'-{\gm'}'\td\varpi^2+\phi_t
-\nu\phi_{\td\varpi\td\varpi}+\sum_{i=1}^md_ie^{a_i^2\kappa
t+a_i\td\varpi}\cos\gm=0\eqno(2.77)$$ by (2.1). We obtain the
following solution of (2.77):
\begin{eqnarray*}\hspace{1cm}\phi&=&\gm'\td\varpi^2+\sum_{s=1}^n\hat
d_se^{\hat a_s^2\kappa t\cos 2\hat b_s+\hat a_s\td\varpi\cos \hat
b_s}\sin(\hat a_s^2\kappa t\sin 2\hat b_s+\hat a_s\td\varpi\sin \hat
b_s+\hat c_s)\\
& &-\sum_{i=1}^md_ie^{a_i^2\nu t+a_i\td\varpi}\int
e^{a_i^2(\kappa-\nu)t}\cos\gm\:dt.\hspace{6.2cm}(2.78)\end{eqnarray*}

Note
$$u=\phi_\varpi\sin\gm-\gm'y,\qquad v=\gm'x
-\phi_\varpi\cos\gm.\eqno(2.79)$$ By (2.72),
\begin{eqnarray*} \hspace{1cm}& & u_t+uu_x+vu_y-\nu\Dlt u\\&=&(\phi_{\varpi
t}-\nu\phi_{\varpi\varpi\varpi})\sin\gm+2\gm'\phi_\varpi\cos\gm-\gm'^2x-{\gm'}'y
\\
&=&(2{\gm'}'\td\varpi-\psi\cos\gm)\sin\gm+2\gm'\phi_\varpi\cos\gm-\gm'^2x-{\gm'}'y,
\\ &=&{\gm'}'(x\sin 2\gm-y\cos 2\gm)
+(2\gm'\phi_\varpi-\psi\sin\gm) \cos\gm-\gm'^2x, \hspace{3cm}(2.80)
\end{eqnarray*}
\begin{eqnarray*} \hspace{1cm}& &v_t+uv_x+vv_y-\nu\Dlt
v-\sta\\ &=&(\nu\phi_{\varpi\varpi\varpi}-\phi_{\varpi
t})\cos\gm+2\gm'\phi_\varpi\sin\gm-\gm'^2y+{\gm'}'x -\psi\\
&=&(\psi\cos\gm-2{\gm'}'\td\varpi)\cos\gm+2\gm'\phi_\varpi\sin\gm-\gm'^2y+{\gm'}'x
-\psi\\ &=&-{\gm'}'(x\cos 2\gm+y\sin
2\gm)+(2\gm'\phi_\varpi-\psi\sin\gm)\sin\gm-\gm'^2y.\hspace{2.8cm}(2.81)
\end{eqnarray*}
 According to (1.1),
$$p=\frac{{\gm'}^2-{\gm'}'\sin 2\gm}{2}x^2+\frac{{\gm'}^2+{\gm'}'\sin
2\gm}{2}y^2+{\gm'}'xy\cos2\gm+\int\psi d\td\varpi\:\sin\gm-2\gm'\phi
\eqno(2.82)$$ modulo the transformation in (1.8)-(1.11).\psp

{\bf Theorem 2.3}. {\it Let $\gm$ be any function of $t$ and denote
$\td\varpi=x\cos\gm+y\sin\gm$. Take
$$\{a_i,b_i,c_i,d_i,\hat a_s,\hat
b_s,\hat c_s,\hat d_s\mid
i=1,...,m;s=1,...,n\}\subset\mbb{R}.\eqno(2.83)$$ If $\nu=\kappa$,
we have the following solutions of the two-dimensional Boussinesq
equations (1.1)-(1.2):
\begin{eqnarray*}u=-\gm' y+\sin\gm\{2\gm'\td\varpi+\sum_{s=1}^n\hat a_s\hat d_se^{\hat a_s^2\kappa
t\cos 2\hat b_s+\hat a_s\td\varpi\cos \hat b_s}\sin(\hat a_s^2\kappa
t\sin 2\hat b_s+\hat a_s\td\varpi\sin \hat b_s+\hat b_s+\hat c_s)\\
-[\sum_{i=1}^m a_id_ie^{a_i^2\kappa t\cos 2b_i+a_i\td\varpi\cos
b_i}\sin(a_i^2\kappa t\sin 2b_i+b_i+a_i\td\varpi\sin b_i+c_i)]\int
\cos\gm\:dt\},\hspace{0.7cm}(2.84)\end{eqnarray*}
\begin{eqnarray*}v=\gm'x-\cos\gm\{2\gm'\td\varpi+\sum_{s=1}^n\hat a_s\hat d_se^{\hat a_s^2\kappa
t\cos 2\hat b_s+\hat a_s\td\varpi\cos \hat b_s}\sin(\hat a_s^2\kappa
t\sin 2\hat b_s+\hat a_s\td\varpi\sin \hat b_s+\hat b_s+\hat
c_s)\\-[\sum_{i=1}^m a_id_ie^{a_i^2\kappa t\cos
2b_i+a_i\td\varpi\cos b_i}\sin(a_i^2\kappa t\sin
2b_i+a_i\td\varpi\sin b_i+b_i+c_i)]\int
\cos\gm\:dt\},\hspace{1cm}(2.85)\end{eqnarray*} $\sta=\psi$ in
(2.73), and
\begin{eqnarray*}
p&=&(\sin\gm+2\gm'\int\cos\gm)[\sum_{i=1}^md_ie^{a_i^2\kappa t\cos
2b_i+a_i\td\varpi\cos b_i}\sin(a_i^2\kappa t\sin
2b_i+a_i\td\varpi\sin b_i+c_i)]\\ & &+\frac{{\gm'}^2-{\gm'}'\sin
2\gm}{2}x^2+\frac{{\gm'}^2+{\gm'}'\sin
2\gm}{2}y^2+{\gm'}'xy\cos2\gm-2\gm'^2\td\varpi^2\\ &
&-2\gm'\sum_{s=1}^n\hat d_se^{\hat a_s^2\kappa t\cos 2\hat b_s+\hat
a_s\td\varpi\cos \hat b_s}\sin(\hat a_s^2\kappa t\sin 2\hat b_s+\hat
a_s\td\varpi\sin \hat b_s+\hat
c_s).\hspace{2.4cm}(2.86)\end{eqnarray*}

When $\nu\neq\kappa$,  we have the following solutions of the
two-dimensional Boussinesq equations (1.1)-(1.2):
\begin{eqnarray*}\hspace{1cm}u&=&\{\sum_{s=1}^n\hat a_s\hat
d_se^{\hat a_s^2\kappa t\cos 2\hat b_s+\hat a_s\td\varpi\cos \hat
b_s}\sin(\hat a_s^2\kappa t\sin 2\hat b_s+\hat a_s\td\varpi\sin \hat
b_s+\hat b_s+\hat c_s)\\
& &+2\gm'\td\varpi-\sum_{i=1}^ma_id_ie^{a_i^2\nu t+a_i\td\varpi}\int
e^{a_i^2(\kappa-\nu)t}\cos\gm\:dt
\}\sin\gm-\gm'y,\hspace{2.3cm}(2.87)\end{eqnarray*}
\begin{eqnarray*}\hspace{1cm}v&=&-\{\sum_{s=1}^n\hat a_s\hat
d_se^{\hat a_s^2\kappa t\cos 2\hat b_s+\hat a_s\td\varpi\cos \hat
b_s}\sin(\hat a_s^2\kappa t\sin 2\hat b_s+\hat a_s\td\varpi\sin \hat
b_s+\hat b_s+\hat c_s)\\
& &+2\gm'\td\varpi-\sum_{i=1}^m a_id_ie^{a_i^2\nu
t+a_i\td\varpi}\int e^{a_i^2(\kappa-\nu)t}\cos\gm\:dt
\}\cos\gm+\gm'x,\hspace{2.3cm}(2.88)\end{eqnarray*} $\sta=\psi$ in
(2.76), and
\begin{eqnarray*}
p&=&\frac{{\gm'}^2-{\gm'}'\sin
2\gm}{2}x^2+\frac{{\gm'}^2+{\gm'}'\sin
2\gm}{2}y^2+{\gm'}'xy\cos2\gm-2\gm'^2\td\varpi^2
\\&&-2\gm'\sum_{s=1}^n\hat
d_se^{\hat a_s^2\kappa t\cos 2\hat b_s+\hat a_s\td\varpi\cos \hat
b_s}\sin(\hat a_s^2\kappa t\sin 2\hat b_s+\hat a_s\td\varpi\sin \hat
b_s+\hat c_s)\\ & &+\sum_{i=1}^m d_ie^{a_i^2\nu
t+a_i\td\varpi}(2\gm'+\sin\gm)\int
e^{a_i^2(\kappa-\nu)t}\cos\gm\:dt).\hspace{5cm}(2.89)\end{eqnarray*}
}\pse

{\bf Remark 2.4}. By Fourier expansion, we can use the above
solution to obtain the one depending on two  piecewise continuous
functions of $\td\varpi$.

\section{Asymmetric Approach I to the 3D  Equations}

Starting from this section, we use  asymmetric approaches developed
in [11] to solve the  stratified rotating Boussinesq equations
(1.3)-(1.7).

For convenience of computation, we denote
$$\Phi_1=u_t+uu_x+vu_y+wu_z-\frac{1}{R_0}v-\sgm(u_{xx}+u_{yy}+u_{zz}),\eqno(3.1)$$
$$\Phi_2=v_t+uv_x+vv_y+wv_z+\frac{1}{R_0}u-\sgm(v_{xx}+v_{yy}+v_{zz}),\eqno(3.2)$$
$$\Phi_3=w_t+uw_x+vw_y+ww_z-\sgm R T-\sgm(w_{xx}+w_{yy}+w_{zz}).\eqno(3.3)$$
Then the equations (1.3)-(1.5) become
 $$\Phi_1+\sgm p_x=0,\qquad
 \Phi_2+\sgm p_y=0,\qquad
\Phi_3+\sgm p_z=0.
 \eqno(3.4)$$
 Our strategy is  to solve the following
 compatibility conditions:
 $$\ptl_y(\Phi_1)=\ptl_x(\Phi_2),\qquad
 \ptl_z(\Phi_1)=\ptl_x(\Phi_3),\qquad\ptl_z(\Phi_2)=\ptl_y(\Phi_3).
 \eqno(3.5)$$

First we assume
$$u=\phi_z(t,z) x+\vs(t,z) y+\mu(t,z),\qquad v=\tau(t,z)
x+\psi_z(t,z) y+\ves(t,z),\eqno(3.6)$$
$$ w=-\phi(t,z)-\psi(t,z),\qquad T=\vt(t,z)+z,\eqno(3.7)$$ where $\phi,\vt,\vs,\mu,\tau,$ and
$\ves$ are functions of $t,z$ to be determined. Then
\begin{eqnarray*}\Phi_1&=&\phi_{tz}x+\vs_t y+\mu_t+
\phi_z(\phi_z x+\vs y+\mu)+(\vs-1/R_0)(\tau x+\psi_zy+\ves)\\ &
&-(\phi+\psi)(\phi_{zz}x+\vs_z y+\mu_z)
-\sgm(\phi_{zzz}x+\vs_{zz} y+\mu_{zz})\\
&=&[\phi_{tz}+\phi_z^2+\tau(\vs-1/R_0)-\phi_{zz}(\phi+\psi)-\sgm\phi_{zzz}]x\\
& &+[\vs_t+\vs\phi_z+\psi_z(\vs-1/R_0)-\vs_z(\phi+\psi)-\sgm \vs_{zz}]y\\
& &+\mu_t+ \mu\phi_z+(\vs-1/R_0)\ves-\mu_z(\phi+\psi)-\sgm\mu_{zz},
\hspace{5.3cm}(3.8)\end{eqnarray*}
\begin{eqnarray*}\Phi_2&=&\tau_tx+\psi_{tz}y+\ves_t+\psi_z(\tau x+\psi_zy+\ves)+
(\tau+1/R_0)(\phi_zx+\vs y+\mu)\\
& &-(\phi+\psi)(\tau_zx+\psi_{zz}y+\ves_z)
-\sgm(\tau_{zz}x+\psi_{zzz}y+\ves_{zz})\\
&=&[\psi_{tz}+\psi_z^2+\vs(\tau+1/R_0)-(\phi+\psi)\psi_{zz}-\sgm\psi_{zzz}]y\\
& &+[\tau_t+\tau\psi_z+(\tau+1/R_0)\phi_z-(\phi+\psi)\tau_z-\sgm
\tau_{zz}]x\\ & &+\ves_t+
\ves\psi_z+(\tau+1/R_0)\mu-(\phi+\psi)\ves_z-\sgm\ves_{zz},
\hspace{5.3cm}(3.9)\end{eqnarray*}
$$\Phi_3=-\phi_t-\psi_t+(\phi+\psi)(\phi_z+\psi_z)-\sgm
R(\vt+z)+\sgm(\phi_{zz}+\psi_{zz}).\eqno(3.10)
$$
Thus (3.5) is equivalent to the following system of partial
differential equations:
$$\phi_{tz}+\phi_z^2+\tau(\vs-1/R_0)-\phi_{zz}(\phi+\psi)-\sgm\phi_{zzz}=\al_1,\eqno(3.11)$$
$$\vs_t+\vs\phi_z+\psi_z(\vs-1/R_0)-\vs_z(\phi+\psi)-\sgm
\vs_{zz}=\al,\eqno(3.12)$$
$$\mu_t+
\mu\phi_z+(\vs-1/R_0)\ves-\mu_z(\phi+\psi)-\sgm\mu_{zz}=\al_2,\eqno(3.13)$$
$$\psi_{tz}+\psi_z^2+\vs(\tau+1/R_0)-(\phi+\psi)\psi_{zz}-\sgm\psi_{zzz}=\be_1,\eqno(3.14)$$
$$\tau_t+\tau\psi_z+(\tau+1/R_0)\phi_z-(\phi+\psi)\tau_z-\sgm
\tau_{zz}=\al,\eqno(3.15)$$
$$\ves_t+\ves\psi_z+(\tau+1/R_0)\mu-(\phi+\psi)\ves_z-\sgm\ves_{zz}=\be_2\eqno(3.16)$$
for some $\al,\al_1,\al_2,\be_1,\be_2$ are functions of $t$.

Let $0\neq b$ and $c$ be fixed real constants.  Recall the notions
in (2.12) and (2.13) with $\gm=b$. We assume
$$\phi=b^{-1}\gm_1\zeta_r(z),\qquad
\psi=b^{-1}(\gm_2\zeta_r(z)+\gm_3\eta_r(z)),\eqno(3.17)$$
$$\vs=\gm_4(\gm_2\eta_r(z)-(-1)^r\gm_3\zeta_r(z)),\qquad\tau=\gm_5\gm_1\eta_r(z),\qquad\gm_4\gm_5=1,\eqno(3.18)$$
where $\gm_i$ are functions of $t$ to be determined. Moreover,
(3.11) becomes
$$(\gm_1'+(-1)^rb^2\sgm\gm_1-\gm_1\gm_5/R_0)\eta_r(z)+(\gm_1+\gm_2)\gm_1c^r
=\al_1,\eqno(3.19)$$ which is implied by
$$\al_1=(\gm_1+\gm_2)\gm_1c^r,\eqno(3.20)$$
$$\gm_1'+(-1)^rb^2\sgm\gm_1-\gm_1\gm_5/R_0=0.\eqno(3.21)$$
On the other hand, (3.15) becomes
$$[(\gm_1\gm_5)'+ \gm_1/R_0+(-1)^rb^2\sgm\gm_1\gm_5]\eta_r+
\gm_1\gm_5(\gm_1+\gm_2)c^r=\al,\eqno(3.22)$$ which gives
$$\al=\gm_1\gm_5(\gm_1+\gm_2)c^r,\eqno(3.23)$$
$$(\gm_1\gm_5)'+(-1)^rb^2\sgm\gm_1\gm_5+ \gm_1/R_0=0.\eqno(3.24)$$ Solving (3.21) and (3.24) for $\gm_1$ and
$\gm_1\gm_5$, we get
$$\gm_1=b_1e^{-(-1)^rb^2\sgm t}\sin\frac{t}{R_0},\qquad\gm_1\gm_5=
b_1e^{-(-1)^rb^2\sgm t}\cos\frac{t}{R_0},\eqno(3.25)$$ where $b_1$
is a real constant. In particular, we take
$$\gm_5=\cot\frac{t}{R_0}.\eqno(3.26)$$

Observe that (3.12) becomes
\begin{eqnarray*}\hspace{1cm}& &[(\gm_2\gm_4)'+(-1)^rb^2\sgm\gm_2\gm_4-\gm_2/R_0]\eta_r(z)
+\gm_4(\gm_1\gm_2+\gm_2^2+(-1)^r\gm_3^2)c^r\\
&&-(-1)^r[(\gm_3\gm_4)'+(-1)^rb^2\sgm\gm_2\gm_4-\gm_3/R_0]\zeta_r(z)=\al
\hspace{4.3cm}(3.27)\end{eqnarray*} and (3.14) becomes
\begin{eqnarray*}\hspace{1cm}& &[\gm_2'+(-1)^rb^2\sgm\gm_2+\gm_2\gm_4/R_0]\eta_r(z)
+(\gm_1\gm_2+\gm_2^2+(-1)^r\gm_3^2)c^r\\
&&-(-1)^r[\gm_3'+(-1)^rb^2\sgm\gm_3+\gm_3\gm_4/R_0]\zeta_r(z)=\be_1,
\hspace{4.8cm}(3.28)\end{eqnarray*} equivalently,
$$\al=\gm_4(\gm_1\gm_2+\gm_2^2+(-1)^r\gm_3^2)c^r,\eqno(3.29)$$
$$\be_1=(\gm_1\gm_2+\gm_2^2+(-1)^r\gm_3^2)c^r,\eqno(3.30)$$
$$(\gm_2\gm_4)'+(-1)^rb^2\sgm\gm_2\gm_4-\gm_2/R_0=0,\eqno(3.31)$$
$$\gm_2'+(-1)^rb^2\sgm\gm_2+\gm_2\gm_4/R_0=0,\eqno(3.32)$$
$$(\gm_3\gm_4)'+(-1)^rb^2\sgm\gm_2\gm_4-\gm_3/R_0=0,\eqno(3.33)$$
$$\gm_3'+(-1)^rb^2\sgm\gm_3+\gm_3\gm_4/R_0=0.\eqno(3.34)$$
Solving (3.31)-(3.34) under the assumption $\gm_4\gm_5=1$, we obtain
$$\gm_2\gm_4=b_2e^{-(-1)^rb^2\sgm t}\sin\frac{t}{R_0},\qquad\gm_2=
b_2e^{-(-1)^rb^2\sgm t}\cos\frac{t}{R_0},\eqno(3.35)$$
$$\gm_3\gm_4=b_3e^{-(-1)^rb^2\sgm t}\sin\frac{t}{R_0},\qquad\gm_3=
b_3e^{-(-1)^rb^2\sgm t}\cos\frac{t}{R_0}.\eqno(3.36)$$ In
particular, we have:
$$\gm_4=\tan\frac{t}{R_0}.\eqno(3.37)$$

According to (3.23) and (3.29),
$$\gm_1\gm_5(\gm_1+\gm_2)c^r=\gm_4(\gm_1\gm_2+\gm_2^2+(-1)^r\gm_3^2)c^r,
\eqno(3.38)$$ equivalently
$$-2b_1b_2\cos\frac{2t}{R_0}+(b_2^2-b_1^2+(-1)^rb_3^2)\sin\frac{2t}{R_0}
=0.\eqno(3.39)$$ Thus
$$b_1b_2=0,\qquad b_2^2-b_1^2+(-1)^rb_3^2=0.\eqno(3.40)$$
So
$$ r=0,\qquad b_2=0,\qquad b_1=b_3\eqno(3.41)$$
or
$$ r=1,\qquad b_1=0,\qquad b_2=b_3.\eqno(3.42)$$

Assume $r=0$ and $b_1\neq 0$. Then
$$\phi=b^{-1}b_1e^{-b^2\sgm t}\sin bz\:\sin\frac{t}{R_0},\qquad \psi=
b^{-1}b_1e^{-b^2\sgm t}\cos bz\:\cos\frac{t}{R_0},\eqno(3.43)$$
$$\vs=-b_1e^{-b^2\sgm t}\sin bz\:\sin\frac{t}{R_0},\qquad\tau=b_1e^{-b^2\sgm
t}\cos bz\:\cos\frac{t}{R_0}.\eqno(3.44)$$ Moreover, we take
$\mu=\ves=\vt=0$. Then
$$\Phi_1=\gm_1^2(x+\gm_5y)=b_1^2e^{-2b^2\sgm
t}\sin\frac{t}{R_0}\left(x\sin\frac{t}{R_0}+y\cos\frac{t}{R_0}\right)\eqno(3.45)$$
by (3.8), (3.11)-(3.12), (3.20) and (3.23). Similarly
$$\Phi_2=b_1^2e^{-2b^2\sgm
t}\cos\frac{t}{R_0}\left(x\sin\frac{t}{R_0}+y\cos\frac{t}{R_0}\right).\eqno(3.46)$$

According to (3.10)
$$\Phi_3=\left[b^{-1}R_0^{-1}b_1e^{-b^2\sgm
t}-b^{-1}b_1^2e^{-2b^2\sgm t}\cos
\left(bz-\frac{t}{R_0}\right)\right]\sin\left(bz-\frac{t}{R_0}\right)-R\sgm
z.\eqno(3.47)$$ By (3.4), we have
\begin{eqnarray*}\hspace{1cm}p&=&\frac{Rz^2}{2}+\frac{b_1e^{-b^2\sgm
t}}{b^2\sgm R_0}\cos
\left(bz-\frac{t}{R_0}\right)-\frac{b_1^2e^{-2b^2\sgm
t}}{2\sgm b^2}\cos ^2\left(bz-\frac{t}{R_0}\right)\\
& &-\frac{b_1^2e^{-2b^2\sgm
t}}{2\sgm}\left(y^2\cos^2\frac{t}{R_0}+x^2\sin^2\frac{t}{R_0}+xy\sin\frac{2t}{R_0}
\right)\hspace{3.7cm}(3.48)\end{eqnarray*} modulo the transformation
in (1.14)-(1.16).

Suppose  $r=1$ and $b_2\neq 0$. Then
$$\phi=\tau=\mu=\ves=\vt=0,\;\;\psi=b^{-1}b_2e^{bz+b^2\sgm t}\cos\frac{t}{R_0},
\qquad\vs=b_2e^{bz+b^2\sgm t}\sin\frac{t}{R_0}.\eqno(3.49)$$
Moreover,
$$\Phi_1=\Phi_2=0,\;\;\Phi_3=b^{-1}b_2R_0^{-1}e^{bz+b^2\sgm
t}\sin\frac{t}{R_0}+b^{-1}b_2^2e^{2(bz+b^2\sgm
t)}\cos^2\frac{t}{R_0}-R\sgm z.\eqno(3.50)$$ According to (3.4),
$$p=\frac{Rz^2}{2}-\frac{b_2e^{bz+b^2\sgm
t}}{b^2\sgm R_0}\sin\frac{t}{R_0}-\frac{b_2^2e^{2(bz+b^2\sgm
t)}}{2b^2\sgm}\cos^2\frac{t}{R_0}\eqno(3.51)$$ modulo the
transformation (1.14)-(1.16).\psp

{\bf Theorem 3.1}. {\it Let $b,b_1,b_2\in\mbb{R}$ with $b\neq 0$. We
have the following solutions of the three-dimensional stratified
rotating Boussinesq equations (1.3)-(1.7): (1)
$$u=b_1e^{-b^2\sgm t}(x\cos bz-y\sin bz)\sin\frac{t}{R_0},\qquad v=
b_1e^{-b^2\sgm t}(x\cos bz-y\sin bz)\cos\frac{t}{R_0},\eqno(3.52)$$
$$w=-b^{-1}b_1e^{-b^2\sgm t}\cos\left(bz-\frac{t}{R_0}\right),\qquad
T=z\eqno(3.53)$$ and $p$ is given in (3.48); (2)
$$u=b_2e^{bz+b^2\sgm t}y\sin\frac{t}{R_0},\qquad v=b_2e^{bz+b^2\sgm
t}y\cos\frac{t}{R_0},\eqno(3.54)$$
$$w=-b^{-1}b_2e^{bz+b^2\sgm
t}\cos\frac{t}{R_0}\qquad T=z\eqno(3.55)$$ and $p$ is given in
(3.51).}\psp

Next we assume $\phi=\vs=\psi=\tau=0$. Then
$$\mu_t-\frac{1}{R_0}\ves-\sgm\mu_{zz}=\al_2,\;\;
\ves_t+\frac{1}{R_0}\nu-\sgm\ves_{zz}=\be_2,\;\;\vt_t-\vt_{zz}=0.\eqno(3.56)$$
Solving them, we get:\psp

{\bf Theorem 3.2}. {\it Let $a_i,b_i,c_i,d_i,\hat a_r,\hat b_r,\hat
c_r,\hat d_r,\td a_s,\td b_s,\td c_s,\td d_s$ be real numbers.
 We have the following solutions of the
three-dimensional stratified rotating Boussinesq equations
(1.3)-(1.7):
\begin{eqnarray*}u&=&\cos\frac{t}{R_0}\;\sum_{i=1}^md_ie^{a_i^2\sgm t\cos
2b_i+ a_iz\cos b_i}\sin(a_i^2\sgm t\sin 2b_i+a_iz\sin b_i+c_i)\\
& &+\sin\frac{t}{R_0}\;\sum_{r=1}^n \hat d_re^{\hat a_r^2\sgm t\cos
2\hat b_r+a_rz\cos\hat b_r}\sin(\hat a_r^2\sgm t\sin 2\hat b_r+\hat
a_rz\sin \hat b_r+\hat c_r),\hspace{1.7cm}(3.57)\end{eqnarray*}
\begin{eqnarray*}v&=&-\sin\frac{t}{R_0}\;\sum_{i=1}^md_ie^{a_i^2\sgm t\cos
2b_i+ a_iz\cos b_i}\sin(a_i^2\sgm t\sin 2b_i+a_iz\sin b_i+c_i)\\
& &+\cos\frac{t}{R_0}\;\sum_{r=1}^n \hat d_re^{\hat a_r^2\sgm t\cos
2\hat b_r+a_rz\cos\hat b_r}\sin(\hat a_r^2\sgm t\sin 2\hat b_r+\hat
a_rz\sin \hat b_r+\hat c_r),\hspace{1.6cm}(3.58)\end{eqnarray*}
$$w=0,\;\;T=z+\sum_{s=1}^k\td a_s\td d_s e^{\td a_s^2 t\cos
2\td b_s+ \td a_sz\cos \td b_s}\sin(\td a_s^2 t\sin 2\td b_s+\td
a_sz\sin \td b_s+\td b_s+\td c_s),\eqno(3.59)$$
$$p=\frac{R z^2}{2}+R\sum_{s=1}^{m_3}\td d_s e^{\td a_s^2 t\cos
2\td b_s+ \td a_sz\cos \td b_s}\sin(\td a_s^2 t\sin 2\td b_s+\td
a_sz\sin \td b_s+\td c_s).\eqno(3.60)$$ }\pse

{\bf Remark 3.3}. By Fourier expansion, we can use the above
solution to obtain the one depending on three arbitrary piecewise
continuous functions of $z$.

\section{Asymmetric Approach II to the 3D  Equations}

In this section, we solve the stratified rotating Boussinesq
equations (1.4)-(1.7) under the assumption
$$u_z=v_z=w_{zz}=T_{zz}=0.\eqno(4.1)$$

Let $\gm$ be a function of $t$ and we use the moving frame
$\td\varpi$ in (2.65). Assume
$$u=f(t,\td\varpi)\sin\gm-\gm'y,\qquad
v=-f(t,\td\varpi)\cos\gm+\gm'x,\eqno(4.2)$$ According to (4.3), we
assume
$$w=\phi(t,\varpi),\qquad
T=\psi(t,\varpi)+z,\eqno(4.3)$$ for some functions $f,\;\phi$ and
$\psi$ in $t$ and $\td\varpi$. Using (2.66)-(2.68), we get
$$\Phi_1=-(\gm'^2+\gm'/R_0)x-{\gm'}'y+f_t\sin\gm+(2\gm'+1/R_0)f\cos\gm-\sgm
f_{\td\varpi\td\varpi}\sin\gm,\eqno(4.4)$$
$$\Phi_2=-(\gm'^2+\gm'/R_0)y+{\gm'}'x-f_t\cos\gm+(2\gm'+1/R_0)f\sin\gm+\sgm
f_{\td\varpi\td\varpi}\cos\gm,\eqno(4.5)$$
$$\Phi_3=\phi_t-\sgm\phi_{\td\varpi\td\varpi}-\sgm
R(\psi+z).\eqno(4.6)$$ By (3.5), we have
$$-2{\gm'}'+f_{\td\varpi t}-\sgm
f_{\td\varpi\td\varpi\td\varpi}=0,\eqno(4.7)$$
$$\phi_t-\sgm\phi_{\td\varpi\td\varpi}-\sgm R\psi=0.\eqno(4.8)$$
Moreover, (1.6) becomes
$$\psi_t-\psi_{\td\varpi\td\varpi}=0.\eqno(4.9)$$

Solving (4.7), we have:
$$f=2\gm'\td\varpi+\sum_{i=1}^m a_id_ie^{a_i^2\kappa t\cos 2b_i+a_i\td\varpi\cos
b_i}\sin(a_i^2\kappa t\sin 2b_i+a_i\td\varpi\sin
b_i+b_i+c_i),\eqno(4.10)$$
 where $a_i,b_i,c_i,d_i$ are arbitrary  real numbers.
Moreover, (4.8) and (4.9) yield
$$\phi=\sum_{r=1}^n \hat d_re^{\hat a_r^2 t\cos 2\hat b_r+\hat a_r\td\varpi\cos
\hat b_r}\sin(\hat a_r^2 t\sin 2\hat b_i+\hat a_r\td\varpi\sin \hat
b_r+\hat c_r)+\sgm Rt\psi,\eqno(4.11)$$
$$\psi=\sum_{s=1}^k\td d_se^{\td a_s^2t\cos 2\td
b_s+\td a_s\td\varpi\cos \td b_s}\sin(\td a_s^2 t\sin 2\td b_s+\td
a_s\td\varpi\sin \td b_s+\td c_s)\eqno(4.12)$$ if $\sgm=1$, and
\begin{eqnarray*}\phi&=&\sum_{r=1}^n \hat d_re^{\hat a_r^2\sgm t\cos 2\hat b_r+\hat a_r\td\varpi\cos
\hat b_r}\sin(\hat a_r^2\sgm t\sin 2\hat b_i+\hat a_r\td\varpi\sin
\hat b_r+\hat c_r)\\ & &+\frac{\sgm R}{1-\sgm}\sum_{s=1}^k\td
d_se^{\td a_s^2t\cos 2\td b_s+\td a_s\td\varpi\cos \td b_s}\sin(\td
a_s^2 t\sin 2\td b_s+\td a_s\td\varpi\sin \td b_s+\td
c_s),\hspace{2.2cm}(4.13)\end{eqnarray*}
$$\psi=\sum_{s=1}^k\td a_s^2\td d_se^{\td a_s^2t\cos 2\td b_s+\td a_s\td\varpi\cos
\td b_s}\sin(\td a_s^2 t\sin 2\td b_s+\td a_s\td\varpi\sin \td
b_s+2\td b_s+\td c_s)\eqno(4.14)$$ when $\sgm\neq 1$, where $\hat
a_r,\hat b_r,\hat c_r,\hat d_r,\td a_s,\td b_s,\td c_s,\td d_s$ are
arbitrary real numbers.

Now
$$\Phi_1=({\gm'}'\sin2\gm-\gm'^2-\gm'/R_0)x-{\gm'}'y\cos2\gm+(2\gm'+1/R_0)f\cos\gm,\eqno(4.15)$$
$$\Phi_2=-({\gm'}'\sin2\gm+\gm'^2+\gm'/R_0)y-{\gm'}'x\cos2\gm+(2\gm'+1/R_0)f\sin\gm\eqno(4.16)$$
and $\Phi_3=-\sgm Rz$. According (3.4), we have
\begin{eqnarray*}p&=&
-\frac{2\gm'+1/R_0}{\sgm}[\gm'\td\varpi^2+\sum_{i=1}^m
d_ie^{a_i^2\kappa t\cos 2b_i+a_i\td\varpi\cos b_i}\sin(a_i^2\kappa
t\sin 2b_i+a_i\td\varpi\sin b_i+c_i)]\\
&&+\frac{R}{2}z^2+\frac{(\gm'^2+\gm'/R_0)(x^2+y^2)+{\gm'}'(y^2-x^2)\sin2\gm}
{2\sgm}+\frac{{\gm'}'}{\sgm}xy\cos2\gm\hspace{1.8cm}(4.17)\end{eqnarray*}
modulo the transformation in (1.14)-(1.16).\psp

{\bf Theorem 4.1}. {\it Let $a_i,b_i,c_i,d_i,\hat a_r,\hat b_r,\hat
c_r,\hat d_r,\td a_s,\td b_s,\td c_s,\td d_s$ be real numbers and
let $\gm$ be any function of $t$. Denote
$\td\varpi=x\cos\gm+y\sin\gm$.
 We have the following solutions of the
three-dimensional stratified rotating Boussinesq equations
(1.3)-(1.7): \begin{eqnarray*}u&=&[\sum_{i=1}^m a_id_ie^{a_i^2\kappa
t\cos 2b_i+a_i\td\varpi\cos b_i}\sin(a_i^2\kappa t\sin
2b_i+a_i\td\varpi\sin b_i+b_i+c_i)\\ & &+2\gm'\td\varpi]\sin\gm-\gm'
y,\hspace{10.1cm}(4.18)\end{eqnarray*}
\begin{eqnarray*}v&=&[-\sum_{i=1}^m a_id_ie^{a_i^2\kappa
t\cos 2b_i+a_i\td\varpi\cos b_i}\sin(a_i^2\kappa t\sin
2b_i+a_i\td\varpi\sin b_i+b_i+c_i)\\ & &+2\gm'\td\varpi]\cos\gm+\gm'
x,\hspace{10.1cm}(4.19)\end{eqnarray*} $p$ is given in (4.17);
\begin{eqnarray*}w&=&\sum_{r=1}^n \hat d_re^{\hat a_r^2 t\cos 2\hat b_r+\hat a_r\td\varpi\cos
\hat b_r}\sin(\hat a_r^2 t\sin 2\hat b_i+\hat a_r\td\varpi\sin \hat
b_r+\hat c_r)\\ & &+\sgm Rt\sum_{s=1}^k\td d_se^{\td a_s^2t\cos 2\td
b_s+\td a_s\td\varpi\cos \td b_s}\sin(\td a_s^2 t\sin 2\td b_s+\td
a_s\td\varpi\sin \td b_s+\td
c_s),\hspace{2.5cm}(4.20)\end{eqnarray*}
$$T=z+\sum_{s=1}^k\td d_se^{\td a_s^2t\cos 2\td
b_s+\td a_s\td\varpi\cos \td b_s}\sin(\td a_s^2 t\sin 2\td b_s+\td
a_s\td\varpi\sin \td b_s+\td c_s)\eqno(4.21)$$ if $\sgm=1$, and
\begin{eqnarray*}w&=&\sum_{r=1}^n \hat d_re^{\hat a_r^2\sgm t\cos 2\hat b_r+\hat a_r\td\varpi\cos
\hat b_r}\sin(\hat a_r^2\sgm t\sin 2\hat b_i+\hat a_r\td\varpi\sin
\hat b_r+\hat c_r)\\ & &+\frac{\sgm R}{1-\sgm}\sum_{s=1}^k\td
d_se^{\td a_s^2t\cos 2\td b_s+\td a_s\td\varpi\cos \td b_s}\sin(\td
a_s^2 t\sin 2\td b_s+\td a_s\td\varpi\sin \td b_s+\td
c_s),\hspace{2.2cm}(4.22)\end{eqnarray*}
$$T=z+\sum_{s=1}^k\td a_s^2\td d_se^{\td a_s^2t\cos 2\td b_s+\td a_s\td\varpi\cos
\td b_s}\sin(\td a_s^2 t\sin 2\td b_s+\td a_s\td\varpi\sin \td
b_s+2\td b_s+\td c_s)\eqno(4.23)$$ when $\sgm\neq 1$. }\psp

{\bf Remark 4.2}. By Fourier expansion, we can use the above
solution to obtain the one depending on three arbitrary piecewise
continuous functions of $\td\varpi$.\psp

Next we let $\al$ be any fixed function of $t$ and set
$$\varpi=\al(x^2+y^2).\eqno(4.24)$$
We assume
$$u=y\phi(t,\varpi)-\frac{\al'}{2\al}x,\qquad
v=-x\phi(t,\varpi)-\frac{\al'}{2\al}y,\eqno(4.25)$$
$$w=\psi(t,\varpi)+\frac{\al'}{\al} z,\qquad T=\vt(t,\varpi)+z\eqno(4.26)$$
where $\phi,\psi$ and $\vt$ are functions in $t,\varpi$. Note
$$\Phi_1=-\frac{{\al'}^2+2\al{\al'}'}{4\al^2}x
+\frac{\al'}{2R_0\al}y+y\phi_t+\left(\frac{x}{R_0}
-\frac{\al'}{\al}y \right)\phi-x\phi^2-4\sgm \al
y(\varpi\phi)_{\varpi \varpi},\eqno(4.27)$$
$$\Phi_2=-\frac{{\al'}^2+2\al{\al'}'}{4\al^2}y
-\frac{\al'}{2R_0\al}x-x\phi_t+\left(\frac{y}{R_0}
+\frac{\al'}{\al}x \right)\phi-y\phi^2+4\sgm \al
x(\varpi\phi)_{\varpi \varpi}.\eqno(4.28)$$
 According to the first equation in (3.5),
$$\left[\varpi\left(\phi_t-\frac{\al'}{\al}\phi-4\sgm \al
(\varpi\phi)_{\varpi
\varpi}\right)\right]_\varpi+\frac{\al'}{2R_0\al}=0,\eqno(4.29)$$
equivalently,
$$(\varpi\phi)_t-\frac{\al'}{\al}\varpi\phi-4\sgm \al
\varpi(\varpi\phi)_{\varpi
\varpi}+\frac{\al'\varpi}{2R_0\al}=\al\be'\eqno(4.30)$$ for some
function $\be$ of $t$. Write
$$\hat\phi=\frac{\varpi\phi}{\al}+\frac{\varpi}{2R_0\al}-\be.\eqno(4.31)$$
Then (4.30) becomes
$$\hat\phi_t-4\sgm
\al\varpi\hat\phi_{\varpi \varpi}=0.\eqno(4.32)$$ Suppose
$$\hat\phi=\sum_{i=1}^\infty\gm_i\varpi^i,\eqno(4.33)$$
where $\gm_i$ are functions of $t$ to be determined. Equation
(4.32) yields
$$(\gm_i)_t=4i(i+1)\sgm\al\gm_{i+1}.\eqno(4.34)$$
Hence
$$\gm_{i+1}=\frac{(\al^{-1}\ptl_t)^i(\gm)}{i!(i+1)!(4\sgm)^i}\eqno(4.35)$$
for some function $\gm$ of $t$. Thus
$$\hat\phi=
\sum_{i=0}^\infty\frac{(\al^{-1}\ptl_t)^i(\gm)\varpi^{i+1}}{i!(i+1)!(4\sgm)^i}.
\eqno(4.36)$$ By (4.31), we get
$$\phi=\frac{\al\be}{\varpi}-\frac{1}{2R_0}+
\al\sum_{i=0}^\infty\frac{(\al^{-1}\ptl_t)^i(\gm)\varpi^i}
{i!(i+1)!(4\sgm)^i}.\eqno(4.37)$$

Note
$$\Phi_3=\psi_t+\frac{\al'}{\al}\psi-4\sgm(\varpi\psi_{\varpi})_{\varpi}-\sgm
R(\vt+z).\eqno(4.38)$$ By the last two equations in (3.5),
$$\psi_t+\frac{\al'}{\al}\psi-4\sgm(\varpi\psi_{\varpi})_{\varpi}-\sgm
R\vt=0\eqno(4.39)$$ modulo the transformation in (1.14)-(1.16). On
the other hand, (1.6) becomes
$$\vt_t-4(\varpi\vt_{\varpi})_{\varpi}=0.\eqno(4.40)$$
Hence
$$\vt=\sum_{i=0}^\infty\frac{\sta_1^{(i)}\varpi^{i+1}}{4^i((i+1)!)^2}\eqno(4.41)$$
modulo the transformation in (1.14)-(1.16), where $\sta_1$ is an
arbitrary function of $t$. Substituting (4.41) into (4.39), we
obtain
$$\psi=
\al^{-1}\sta_2\varpi+ \al^{-1}\sum_{i=1}^\infty\frac{\sta_2^{(i)}+
R\sum_{r=0}^{i-1}\sgm^{i-r}(\al\sta_1^{(i-s-1)})^{(s)}}{(4\sgm)^i((i+1)!)^2}
\varpi^{i+1},\eqno(4.42)$$ where $\sta_2$ is another arbitrary
function of $t$.

Now
$$\Phi_1=-\frac{{\al'}^2+2\al{\al'}'}{4\al^2}x
+\frac{\al\be' y}{\varpi}+\frac{x}{R_0}\phi-x\phi^2,\eqno(4.43)$$
$$\Phi_2=-\frac{{\al'}^2+2\al{\al'}'}{4\al^2}y
-\frac{\al\be' x}{\varpi}+\frac{y}{R_0}\phi-y\phi^2\eqno(4.44)$$ by
(4.27) and (4.28), and
$$\Phi_3=(\al^{-1}\al'-\sgm R)z\eqno(4.45)$$
by (4.38). According to (3.4), we obtain
\begin{eqnarray*}p&=&\left(\frac{{\al'}^2+2\al{\al'}'}{4\sgm\al^2}+\frac{3}{8\sgm R_0^2}\right)(x^2+y^2)
+\frac{\be'}{\sgm}\arctan\frac{y}{x} +\frac{(R_0\al\gm-1)\be}{\sgm
R_0}\ln\al(x^2+y^2)\\ &
&-\frac{\sgm^{-1}\be^2}{2(x^2+y^2)}+\frac{\sgm
R-\al^{-1}\al'R}{2\sgm}z^2 -\frac{1}{\sgm
R_0}\sum_{i=0}^\infty\frac{(\al^{-1}\ptl_t)^i(\gm)\al^{i+1}(x^2+y^2)^{i+1}}
{((i+1)!)^2(4\sgm)^i}\\ & &+\frac{\al}{2\sgm}\sum_{i,j=0}^\infty
\frac{(\al^{-1}\ptl_t)^i(\gm)(\al^{-1}\ptl_t)^j(\gm)(\al(x^2+y^2))^{i+j+1}}
{i!j!(i+1)!(j+1)!(i+j+1)(4\sgm)^{i+j}}\\ &
&+\frac{\al\be}{2\sgm}\sum_{i=1}^\infty\frac{(\al^{-1}\ptl_t)^i(\gm)(\al(x^2+y^2))^i}
{i!(i+1)!i(4\sgm)^i}\hspace{7.6cm}(4.46)
\end{eqnarray*}
modulo the transformation in (1.14)-(1.16). By (4.25), (4.26),
(4.37), (4.41) and (4.42), we have:\psp

{\bf Theorem 4.3} {\it Let $\al,\be,\gm,\sta_1,\sta_2$ be any
function of $t$ such that the following involved series converge. We
have the following solutions of the three-dimensional stratified
rotating Boussinesq equations (1.3)-(1.7):
$$u=\frac{\be y}{x^2+y^2}-\frac{y}{2R_0}-\frac{\al'}{2\al}x+
\al y\sum_{i=0}^\infty\frac{(\al^{-1}\ptl_t)^i(\gm)\al^i(x^2+y^2)^i}
{i!(i+1)!(4\sgm)^i},\eqno(4.47)$$
$$v=\frac{x}{2R_0}-\frac{\al'}{2\al}y-\frac{\be x}{x^2+y^2}+
\al x\sum_{i=0}^\infty\frac{(\al^{-1}\ptl_t)^i(\gm)\al^i(x^2+y^2)^i}
{i!(i+1)!(4\sgm)^i},\eqno(4.48)$$
$$w=\sta_2(x^2+y^2)+\frac{\al'}{\al} z+\frac{1}{\al}\sum_{i=1}^\infty\frac{\sta_2^{(i)}+
R\sum_{r=0}^{i-1}\sgm^{i-r}(\al\sta_1^{(i-s-1)})^{(s)}}{(4\sgm)^i((i+1)!)^2}
\al^{i+1}(x^2+y^2)^{i+1},\eqno(4.49)$$
$$T=z+\sum_{i=0}^\infty\frac{\sta_1^{(i)}\al^{i+1}(x^2+y^2)^{i+1}}{4^i((i+1)!)^2}
\eqno(4.50)$$ and $p$ is given in (4.46).}

\section{Asymmetric Approach III to the 3D  Equations}

In this section, we solve (1.3)-(1.7) with $v_x=w_x=T_x=0$.

Let $c$ be a real constant. Set
$$\varpi=y\cos c+z\sin c.\eqno(5.1)$$
Suppose
$$u=f(t,\varpi),\qquad
v=\phi(t,\varpi)\sin c,\eqno(5.2)$$
$$w=-\phi(t,\varpi)\cos c,\qquad
T=\psi(t,\varpi)+z,\eqno(5.3)$$ where $f,\;\phi$ and $\psi$ are
functions in $t$ and $\varpi$. Then
$$\Phi_1=f_t-\sgm
f_{\varpi\varpi}-\frac{\sin c}{R_0}\phi,\eqno(5.4)$$
$$\Phi_2=(\phi_t-\sgm\phi_{\varpi\varpi})\sin c+\frac{1}{R_0}f,\eqno(5.5)$$
$$\Phi_3=(\sgm\phi_{\varpi\varpi}-\phi_t)\cos c-\sgm R(\psi+z).\eqno(5.6)$$ By (3.5),
$$f_{\varpi t}-\sgm
f_{\varpi\varpi\varpi}-\frac{\sin
c}{R_0}\phi_{\varpi}=0,\eqno(5.7)$$
$$(\phi_t-\sgm\phi_{\varpi\varpi})_{\varpi}+\frac{\sin c}{R_0}f_{\varpi}+\sgm
R\psi_{\varpi}\cos c=0.\eqno(5.8)$$ Modulo (1.14)-(1.16), we have
$$f_t-\sgm
f_{\varpi\varpi}-\frac{\sin c}{R_0}\phi=0,\eqno(5.9)$$
$$\phi_t-\sgm\phi_{\varpi\varpi}+\frac{\sin c}{R_0}f+\sgm
R\psi\cos c=0.\eqno(5.10)$$

Denote
$$\left(\begin{array}{c}\hat
f\\\hat\phi\end{array}\right)=\left(\begin{array}{cc}\cos\frac{t\sin
c}{R_0}&-\sin\frac{t\sin c}{R_0}\\ \sin\frac{t\sin
c}{R_0}&\cos\frac{t\sin
c}{R_0}\end{array}\right)\left(\begin{array}{c}
f\\\phi\end{array}\right).\eqno(5.11)$$ Then (5.9) and (5.10) become
$$\hat f_t-\sgm\hat f_{\varpi\varpi}-\sgm R\psi\cos c\;\sin\frac{t\sin
c}{R_0}=0,\eqno(5.12)$$
$$\hat\phi_t-\sgm\hat\phi_{\varpi\varpi}+\sgm R\psi\cos c\;\cos\frac{t\sin
c}{R_0}=0.\eqno(5.13)$$ On the other hand, (1.6) becomes
$$\psi_t-\psi_{\varpi\varpi}=0.\eqno(5.14)$$

Assume $\sgm=1$. We have the following solution:
$$\psi=\sum_{i=1}^m a_id_ie^{a_i^2t\cos 2b_i+a_i\varpi\cos
b_i}\sin(a_i^2t\sin 2b_i+a_i\varpi\sin b_i+b_i+c_i),\eqno(5.15)$$
\begin{eqnarray*}\hat f&=&- RR_0\cot c\;\cos\frac{t\sin
c}{R_0}\;\sum_{i=1}^m a_id_ie^{a_i^2 t\cos 2b_i+a_i\varpi\cos
b_i}\sin(a_i^2t\sin 2b_i+a_i\varpi\sin b_i+b_i+c_i)\\ &
&+\sum_{r=1}^n \hat a_r\hat d_re^{\hat a_r^2 t\cos 2\hat b_r+\hat
a_r\varpi\cos \hat b_r}\sin(\hat a_r^2t\sin 2\hat b_i+\hat
a_r\varpi\sin \hat b_r+\hat b_r+\hat
c_r),\hspace{2.1cm}(5.16)\end{eqnarray*}
\begin{eqnarray*}\hat \phi&=&- RR_0\cot c\;\sin\frac{t\sin
c}{R_0}\;\sum_{i=1}^m a_id_ie^{a_i^2t\cos 2b_i+a_i\varpi\cos
b_i}\sin(a_i^2t\sin 2b_i+a_i\varpi\sin b_i+b_i+c_i)\\ &
&+\sum_{s=1}^k\td a_s\td d_se^{\td a_s^2t\cos 2\td b_s+\td
a_s\varpi\cos \td b_s}\sin(\td a_s^2 t\sin 2\td b_s+\td
a_s\varpi\sin \td b_s+\td b_s+\td
c_s),\hspace{2.1cm}(5.17)\end{eqnarray*} where $a_i,b_i,c_i,\hat
a_r,\hat b_r,\hat c_r,\hat d_r, \td a_s,\td b_s,\td c_s,\td d_s$
 are arbitrary  real numbers. According to (5.11),
\begin{eqnarray*}f=-RR_0\cot c\;\cos\frac{2t\sin
c}{R_0}\;\sum_{i=1}^m a_id_ie^{a_i^2t\cos 2b_i+a_i\varpi\cos
b_i}\sin(a_i^2t\sin 2b_i+a_i\varpi\sin b_i+b_i+c_i)
\\ +\cos\frac{t\sin c}{R_0}\;\sum_{r=1}^n \hat a_r\hat d_re^{\hat a_r^2 t\cos 2\hat b_r+\hat
a_r\varpi\cos \hat b_r}\sin(\hat a_r^2t\sin 2\hat b_i+\hat
a_r\varpi\sin \hat b_r+\hat b_r+\hat c_r)\hspace{2.6cm} \\
+\sin\frac{t\sin c}{R_0}\;\sum_{s=1}^k\td a_s\td d_se^{\td
a_s^2t\cos 2\td b_s+\td a_s\varpi\cos \td b_s}\sin(\td a_s^2 t\sin
2\td b_s+\td a_s\varpi\sin \td b_s+\td b_s+\td
c_s),\hspace{1.5cm}(5.18)\end{eqnarray*}
\begin{eqnarray*}& &\phi=-\sin\frac{t\sin c}{R_0}\;\sum_{r=1}^n \hat a_r\hat d_re^{\hat a_r^2 t\cos 2\hat b_r+\hat
a_r\varpi\cos \hat b_r}\sin(\hat a_r^2t\sin 2\hat b_i+\hat
a_r\varpi\sin \hat b_r+\hat b_r+\hat c_r)\\ & &+\cos\frac{t\sin
c}{R_0}\;\sum_{s=1}^k\td a_s\td d_se^{\td a_s^2t\cos 2\td b_s+\td
a_s\varpi\cos \td b_s}\sin(\td a_s^2 t\sin 2\td b_s+\td
a_s\varpi\sin \td b_s+\td b_s+\td
c_s).\hspace{0.8cm}(5.19)\end{eqnarray*}

Suppose $\sgm\neq 1$. We take the following solution of
(5.11)-(5.14):
$$\psi=\sum_{i=1}^m a_id_ie^{a_i^2t+a_i\varpi},\eqno(5.20)$$
\begin{eqnarray*}\hat f&=&\sgm R\sum_{i=1}^m
a_id_ie^{a_i^2t+a_i\varpi}\frac{\cos
c\:\left[a_i^2(1-\sgm)\sin\frac{t\sin c}{R_0}-R_0^{-1}\sin
c\:\cos\frac{t\sin c}{R_0}\right]}{a_i^4(1-\sgm)^2+R_0^{-2}\sin^2c}
\\ & &+\sum_{r=1}^n \hat a_r\hat d_re^{\hat a_r^2\sgm t\cos
2\hat b_r+\hat a_r\varpi\cos \hat b_r}\sin(\hat a_r^2\sgm t\sin
2\hat b_i+\hat a_r\varpi\sin \hat b_r+\hat b_r+\hat
c_r),\hspace{1.5cm}(5.21)\end{eqnarray*}
\begin{eqnarray*}\hat \phi&=&\sgm R\sum_{i=1}^m
a_id_ie^{a_i^2t+a_i\varpi}\frac{\cos
c\:\left[a_i^2(\sgm-1)\cos\frac{t\sin c}{R_0}-R_0^{-1}\sin
c\:\sin\frac{t\sin c}{R_0}\right]}{a_i^4(1-\sgm)^2+R_0^{-2}\sin^2c}
\\ &
&+\sum_{s=1}^k\td a_s\td d_se^{\td a_s^2\sgm t\cos 2\td b_s+\td
a_s\varpi\cos \td b_s}\sin(\td a_s^2\sgm t\sin 2\td b_s+\td
a_s\varpi\sin \td b_s+\td b_s+\td
c_s),\hspace{1.4cm}(5.22)\end{eqnarray*} where $a_i,b_i,c_i,\hat
a_r,\hat b_r,\hat c_r,\hat d_r, \td a_s,\td b_s,\td c_s,\td d_s$
 are arbitrary  real numbers. According to (5.11),
\begin{eqnarray*}f&=&\cos\frac{t\sin c}{R_0}\;\sum_{r=1}^n \hat a_r\hat d_re^{\hat a_r^2\sgm t\cos 2\hat b_r+\hat
a_r\varpi\cos \hat b_r}\sin(\hat a_r^2\sgm t\sin 2\hat b_i+\hat
a_r\varpi\sin \hat b_r+\hat b_r+\hat c_r) \\&& +\sin\frac{t\sin
c}{R_0}\;\sum_{s=1}^k\td a_s\td d_se^{\td a_s^2\sgm t\cos 2\td
b_s+\td a_s\varpi\cos \td b_s}\sin(\td a_s^2\sgm t\sin 2\td b_s+\td
a_s\varpi\sin \td b_s+\td b_s+\td c_s)\\
& &-\sgm R\sum_{i=1}^m\frac{a_id_ie^{a_i^2t+a_i\varpi}\sin
2c}{2R_0(a_i^4(1-\sgm)^2+R_0^{-2}\sin^2c)}
,\hspace{6.6cm}(5.23)\end{eqnarray*}
\begin{eqnarray*}\phi&=&-\sin\frac{t\sin c}{R_0}\;\sum_{r=1}^n \hat a_r\hat d_re^{\hat a_r^2\sgm t\cos 2\hat b_r+\hat
a_r\varpi\cos \hat b_r}\sin(\hat a_r^2\sgm t\sin 2\hat b_i+\hat
a_r\varpi\sin \hat b_r+\hat b_r+\hat c_r)\\ & &+\cos\frac{t\sin
c}{R_0}\;\sum_{s=1}^k\td a_s\td d_se^{\td a_s^2\sgm t\cos 2\td
b_s+\td a_s\varpi\cos \td b_s}\sin(\td a_s^2\sgm t\sin 2\td b_s+\td
a_s\varpi\sin \td b_s+\td b_s+\td
c_s)\\
& &-\sgm R\sum_{i=1}^m\frac{a_i^3d_i(\sgm-1)e^{a_i^2t+a_i\varpi}\cos
c}{a_i^4(1-\sgm)^2+R_0^{-2}\sin^2c}
.\hspace{7.5cm}(5.24)\end{eqnarray*}

By (5.4)-(5.6), (5.9) and (5.10), $\Phi_1=0$,
$$\Phi_2=\left(\frac{\cos c}{R_0}f-\sgm R\psi\sin c\right)\cos
c,\eqno(5.25)$$
$$\Phi_3=\left(\frac{\cos c}{R_0}f-\sgm R\psi\sin c\right)\sin
c-\sgm Rz.\eqno(5.26)$$ According to (3.4),
\begin{eqnarray*}p&=&\frac{R\cos^2 c}{\sin c}\cos\frac{2t\sin
c}{R_0}\;\sum_{i=1}^m d_ie^{a_i^2t\cos 2b_i+a_i\varpi\cos
b_i}\sin(a_i^2t\sin 2b_i+a_i\varpi\sin b_i+c_i)
\\ & &-\frac{\cos c}{ R_0}\cos\frac{t\sin c}{R_0}\;\sum_{r=1}^n \hat d_re^{\hat a_r^2 t\cos 2\hat b_r+\hat
a_r\varpi\cos \hat b_r}\sin(\hat a_r^2t\sin 2\hat b_i+\hat
a_r\varpi\sin \hat b_r+\hat c_r) \\&& -\frac{\cos c}{
R_0}\sin\frac{t\sin c}{R_0}\;\sum_{s=1}^k\td d_se^{\td a_s^2t\cos
2\td b_s+\td a_s\varpi\cos \td b_s}\sin(\td a_s^2 t\sin 2\td b_s+\td
a_s\varpi\sin \td b_s+\td c_s)\\ & &+R\sin c\:\sum_{i=1}^m
d_ie^{a_i^2t\cos 2b_i+a_i\varpi\cos b_i}\sin(a_i^2t\sin
2b_i+a_i\varpi\sin
b_i+c_i)+\frac{R}{2}z^2\hspace{1.2cm}(5.27)\end{eqnarray*} modulo
 if $\sgm=1$, and
\begin{eqnarray*}p&=&-\frac{\cos c}{\sgm R_0}\cos\frac{t\sin c}{R_0}\;\sum_{r=1}^n \hat d_re^{\hat a_r^2\sgm t\cos 2\hat b_r+\hat
a_r\varpi\cos \hat b_r}\sin(\hat a_r^2\sgm t\sin 2\hat b_i+\hat
a_r\varpi\sin \hat b_r+\hat c_r) \\&& -\frac{\cos c}{\sgm
R_0}\sin\frac{t\sin c}{R_0}\;\sum_{s=1}^k\td d_se^{\td a_s^2\sgm
t\cos 2\td b_s+\td a_s\varpi\cos \td b_s}\sin(\td a_s^2\sgm t\sin
2\td b_s+\td
a_s\varpi\sin \td b_s+\td c_s)\\
& &+ \sum_{i=1}^m\frac{d_iRe^{a_i^2t+a_i\varpi}\sin 2c\;\cos
c}{2R_0^2(a_i^4(1-\sgm)^2+R_0^{-2}\sin^2c)}+R\sin c\;\sum_{i=1}^m
d_ie^{a_i^2t+a_i\varpi}+\frac{R}{2}z^2
,\hspace{1.7cm}(5.28)\end{eqnarray*}
 modulo the transformation in
(1.14)-(1.16).

In summary, we have:\psp

{\bf Theorem 5.1}. {\it Let $a_i,b_i,c_i,\hat a_r,\hat b_r,\hat
c_r,\hat d_r, \td a_s,\td b_s,\td c_s,\td d_s,c$ be arbitrary real
numbers. Denote $\varpi=y\cos x+z\sin c$. We have the following
solutions of the three-dimensional stratified rotating Boussinesq
equations (1.3)-(1.7):
$$u=f,\qquad
v=\phi\sin c,\qquad w=-\phi\cos c,\qquad T=\psi+z,\eqno(5.29)$$
where (1) $f$ is given in (5.18), $\phi$ is given in (5.19), $\psi$
is given in (5.15) and $p$ is given in (5.27) if $\sgm=1$; (2) $f$
is given in (5.23), $\phi$ is given in (5.24), $\psi$ is given in
(5.20) and $p$ is given in (5.28) when $\sgm\neq 1$.} \psp

{\bf Remark 5.2}. By Fourier expansion, we can use the above
solution to obtain the one depending on three arbitrary piecewise
continuous functions of $\varpi$. Applying the transformation ${\cal
T}_1$ in (1.12)-(1.13) to the above solution, we get a solution
involving all the variables $t,x,y,z$.

\bibliographystyle{amsplain}

\begin{thebibliography}{10}



\bibitem{} D. Chae, Global regularity for the 2D Boussinesq
equations with partial viscosity, {\it Adv. Math.} {\bf 203} (2006),
497-513.

\bibitem{} M. Gill and S. Childress, {\it Topics in Geophysical
Fluid Dynamics, Atmospheric Dynamics, Dynamo Theory, and Climate
Dynamics}, Springer-verlag, New York, 1987.

\bibitem{} T. Hou and C. Li, Global well-posedness of the viscous Boussinesq
equations, {\it Discrete Contin. Dyn. Syst.} {\bf 12} (2005), 1-12.

\bibitem{} C. Hsia, T. Ma and S. Wang, Stratified rotating Boussinesq
equations in geophysical fluid dynamics: dynamic bifurcation and
periodic solutions, {\it J. Math. Phys.} {\bf 48} (2007), no. 6,
06560.

\bibitem{} E. N. Lorenz, Deterministic nonperiodic flow, {\it J.
Atmos. Sci.} {\bf 20} (1963), 130-141.

\bibitem{} J. Lions, R. Teman and S. Wang, New formulations of the
primitive equations of the atmosphere and applications, {\it
Nonlinearity} {\bf 5} (1992), 237-288.

\bibitem{} J. Lions, R. Teman and S. Wang, On the equations of
large-scale ocean, {\it Nonlinearity} {\bf 5} (1992), 1007-1053.

\bibitem{} A. Majda, {\it Introduction to PDEs and Waves for the
Atmosphere and Ocean}, Courant Lecture Note in Mathematics, Vol. 9,
AMS and CIMS, 2003.

\bibitem{} J. Pedlosky, {\it Geophsical Fluid Dynamics}, 2rd
Edition, Springer-verlag, New York, 1987.

\bibitem{}  X. Xu,  Stable-Range approach to the equation of nonstationary
transonic gas flows, {\it Quart. Appl. Math.} {\bf 65} (2007),
529-547.

\bibitem{}  X. Xu, Asymmetric and moving-frame  approaches to Navier-Stokes
equations, {\it  Quart. Appl. Math.}, in press, arXiv:0706.1861.

\end{thebibliography}

\end{document}